\documentclass[%
 aip,
 amsmath,amssymb,
 reprint,%
]{revtex4-1}

\usepackage{graphicx}
\usepackage{dcolumn}
\usepackage{bm}

\usepackage[utf8]{inputenc}
\usepackage[T1]{fontenc}
\usepackage{mathptmx}
\usepackage{etoolbox}
\usepackage{bm}
\usepackage{subfig}
\usepackage{stackengine}

\makeatletter
\def\@email#1#2{%
 \endgroup
 \patchcmd{\titleblock@produce}
  {\frontmatter@RRAPformat}
  {\frontmatter@RRAPformat{\produce@RRAP{*#1\href{mailto:#2}{#2}}}\frontmatter@RRAPformat}
  {}{}
}%
\makeatother

\begin{document}

\title[]{Deformations of an active liquid droplet}
\author{R. Kree}
\email{kree@theorie.physik.uni-goettingen.de}
\author{ A. Zippelius}%

\affiliation{ 
Institut f. Theoretische Physik, Universit\"at G\"ottingen, Friedrich-Hund Pl. 1, 37077 G\"ottingen, Germany
}%

\date{\today}

\begin{abstract}
  A fluid droplet in general deforms, if subject to active driving,
  such as a finite slip velocity or active tractions on its
  interface. We show that these deformations and their dynamics can be
  computed analytically in a perturbation theory in the inverse
  surface tension ($\gamma$) using an approach based on vector
  spherical harmonics. In lowest order, the deformation is of order
  $\epsilon\propto 1/\gamma$, yet it affects the flow fields inside
  and outside of the droplet in order $\epsilon^0$. Hence a correct
  description of the flow has to allow for shape fluctuations, -- even
  in the limit of large surface tension.
  
  \end{abstract}
\pacs{47. 63. Gd, 87. 17. Jj, 87. 85. Tu}

\maketitle


\date{\today}


\section{Introduction}
Droplets which are suspended in an ambient fluid are easily
deformed, e.g. by external forces 
or shear flow in the ambient fluid.  Building on early work by
Taylor\cite{Taylor1934}, deformations
 of a passive droplet in shear flow have been
analysed\cite{Kennedy,Acrivos,Rallison1984,Cox,Narsimhan,Vlahovska2009},
assuming inertial effects to be negligible, justified by the small
size of the droplet and the correspondingly low Reynolds
number. Furthermore a large surface tension $\gamma$ is assumed in
order to keep the deformations from spherical shape small, so that they can be studied in a perturbative approach in $\epsilon\propto 1/\gamma$.

Recently, active fluid droplets have moved into focus, mainly because
of their relevance for biological and medical systems and applications. Droplets driven chemically
were shown to grow and divide in a way which is reminiscent of living
cells~\cite{Zwicker2}. Artificial cells are synthesized, starting from
liquid droplets~\cite{Spatz2017}. Living microorganisms are capable of
self propulsion by a variety of mechanims~\cite{Lauga_2009}, one of
them involving controlled shape changes. Outside the living world,
liquid droplets are known to self propel by phoretic
effects\cite{Golestanian2007,Herminghaus2012,Herminghaus2013,Maass2016}; the best
known one is based on an inhomogeneous surface tension -- the
Marangoni effect~\cite{Lauga2012a,Stark_EPL,Schmitt2016}. The flow fields, both inside and outside the
microorganism or artificial swimmer, have been measured. Theoretical
approaches have mainly focused on droplets of spherical shape, using Stokes equations to analyse the propulsion velocities as well as the
flows~\cite{Lauga_2009,Schmitt2016,Kree_2017}. Deformations away from the spherical shape have been assumed to be completely suppressed by a large homogeneous interfacial tension.  Deformable active droplets have been discussed near the onset of the Marangoni instability~\cite{Morozov2019}. Restricting deformations to ellipsoidal shape of the droplet, coupled equations
of motion were derived for the nematic order parameter and the
velocity of a 2-dimensional deformable drop in a quiescent solvent~\cite{Ohta2009} as well as in shear flow~\cite{Tarama2013}.

Here we consider active droplets, driven by either an active slip
velocity or by active tractions on the interface.  These two types of
drives comprise the vast majority of hydrodynamic models of
self-propulsion at small scales. In addition to the intrinsic motion,
such drives also lead to deformations of the droplet, which we focus
on in the present work. We compute the dynamics of deformations to
first order of regular perturbation theory in
$\epsilon\propto 1/\gamma$, using a versatile analytic framework based
on vector spherical harmonics. The boundary value problem stays
linear, and the calculation of propulsion velocities is completely
decoupled from that of deformations. We show that deformations of
$\mathcal{O}(\epsilon)$ contribute to the flow fields in
${\cal{O}}(\epsilon^0)$ , so that even in the limit of
infinite interfacial tension, deformations still influence the flow
fields, although they do no longer modify the spherical shape.

Our results also allow us to quantify errors in the flow fields for stationary deformations, which are computed from  non-deformation boundary conditions~\cite{Schmitt2016,Lauga2017}. These boundary conditions assume a persistent spherical shape of the drop, but do only balance tangential forces at the interface, whereas radial force components are not considered. Even in the limit of infinite homogeneous interfacial tension the results from non-deformation boundary conditions differ significantly from the leading order of perturbation theory. However,  the errors do not show up in the calculation of propulsion velocities, because the corresponding  part of the flow does not deform the droplet.

The model of the droplet and its intrinsic drives is introduced in the
next section, and the perturbative approach is outlined in
Sect.\ref{perturbation}. This is followed in Sect.\ref{sec:analytical}
by a discussion of the analytical solution for the flow field and the
dynamics of deformations. We introduce the general method and consider
a simple example before we discuss the full solution.
Finally, in
Sect.\ref{sect:Comparison}  we specialize
to stationary deformations, which are calculated to
$\mathcal{O}(\epsilon)$, together with the corresponding flow fields.
To illustrate our approach, we discuss a few special cases, including an inhomogeneous surface tension. The flow fields are compared to
flows computed from non-deformation boundary conditions and the
discrepancies are discussed. Although the calculation of propulsion
velocities generated by the drives is not the main focus of this work,
we have included it in Appendix\ref{appendixB} to show that our
framework leads to the same well-known results as the application of
non-deformation boundary conditions.
Details of the derivation of the general solution are given in
Appendix\ref{appendixA}.


\section{Model}
\label{model}
We consider a droplet, consisting of an incompressible Newtonian fluid
with viscosity $\eta^-$. It is immersed into an ambient Newtonian
fluid of viscosity $\eta^+$, which is at rest in the laboratory frame
(LF) at initial time $t=0$. The droplet shape at time $t$ is determined by a level set function $H(\bm{r},t)=0$ so that its outward normal unit vector is $\bm{n}=\nabla H/|\nabla H|$.  
The two fluids are assumed to be completely immiscible, and the droplet is considered neutrally buoyant. In the absence of any driving, the shape of the droplet is spherical  and its radius is $R$. We choose units of mass, length and
time such that the density $\rho_0=1$, the typical size of the droplet $R=1$ and the viscosity of the exterior fluid $\eta^+=1$.
 We do, however, keep the notation $\eta^+$, because some results are more intuitive in the explicit notation.  
 
For low Reynolds number, the flow fields inside ($\bm{v}^-$) and
outside ($\bm{v}^+$) of the droplet are calculated from Stokes's
equation
\begin{equation}
\label{eq:stokes}
\nabla\cdot\bm{\sigma}^{\pm}=\eta^{\pm}\nabla^2\bm{v}^{\pm}-\nabla p^{\pm}=0 , 
\end{equation}
supplemented by the incompressibility condition $\nabla\cdot\bm{v}^{\pm}=0$.
The viscous stress tensor $\bm{\sigma}^{\pm}$ is given by its
cartesian components
$\sigma_{ij}^{\pm}=-p^{\pm}\delta_{ij}+\eta^{\pm}(\partial_iv_j^{\pm}+\partial_jv_i^{\pm})$, with
the pressure $p$ determined from incompressibility.

\subsection{Boundary conditions}
The initially spherical drop is driven by active tangential slip velocities $\bm{v}_{act}$ and/or active tractions $\bm{t}_{act}$,  located at the interface.  Both types of drives are typical mechanisms of many natural and artificial microswimmers. Active tangential slip has become a standard model of self-propelled phoretic particles \cite{Anderson1989} and cilia-driven microorganisms \cite{Blake_1971,Lighthill_1952}. Active tractions are generated, e.g., by the Marangoni effect on the surface of droplets\cite{} \cite{}. In the following we restrict the discussion to axisymmetric and achiral systems. 
The velocities  $\bm{v}_{act}(\theta)=v_\theta(\theta)\bm{e}_\theta$  and the active tractions $\bm{t}_{act}=t_r(\theta)\bm{e}_r + t_\theta(\theta)\bm{e}_\theta$
are represented in standard spherical coordinates $r, \theta, \phi$ with  corresponding unit vectors $\bm{e}_r, \bm{e}_\theta, \bm{e}_\phi$.
They do not depend on the $\phi$-coordinate and we do not allow for components in the $\bm{e}_\phi$ direction, so that  the boundary value problem is reduced to two dimensions. 

The boundary  condition for the flow fields on the interface is given by 
\begin{equation}
  \label{eq:bc2}
\bm{v}^+(\bm{r})=\bm{v}^-(\bm{r})+\bm{v}_{act}. 
\end{equation}
 Note that the active slip velocity is purely tangential, so that the radial components are continuous across the interface.
 
The balance of forces at the interface depends upon the shape of the droplet and takes on the form
\begin{equation}
  \label{eq:bc3}
\bm{t}^+-\bm{t}^-=\gamma\bm{n}(\bm{\nabla}\cdot\bm{n})+\bm{t}_{act},
\end{equation}
with viscous tractions
$\bm{t}^{\pm}=\bm{\sigma}^{\pm}\cdot\bm{n}$. For homogeneous $\gamma$,
the first term on the right hand side corresponds to the Laplace
pressure. It will limit and restore deformations away from a spherical
shape. The last term represents active tractions.  If the surface
tension is inhomogeneous, $\gamma(\theta)=\gamma+\delta\gamma(\theta)$, there will be additional tractions, which are counted among the active drives, and they contribute 
\begin{equation}\label{Marangoni}
  \bm{t}_{act} =2\delta\gamma(\theta)\bm{e}_r -
  \delta\gamma'(\theta)\bm{e}_\theta
      \end{equation}
on a spherical interface.

The six equations Eqs.(\ref{eq:bc2},\ref{eq:bc3}) are
sufficient to determine a unique solution of the Stokes equation for
the initial sphere.  
The calculated flow $\bm{v}^\pm$ will, however, lead to a deformation of the spherical droplet, whenever $\bm{v}^\pm\cdot\bm{e}_r$ does not vanish at $r=1$. We describe the shape evolution by a level set function 
 $H(\bm{r},t)=0$ for all $t$, which implies the kinematic equation
\begin{equation}
\label{eq:kinetics}
  \frac{dH}{dt}=0=\frac{\partial H}{\partial t}+\bm{v}^\pm\cdot\nabla H.
\end{equation}
Here, only the continuous normal component $\propto \bm{v}^\pm\cdot\nabla H$ of the flow velocity on the interface enters.  
This equation determines $H$, provided it is possible to solve the boundary value problem for all possible shapes, --- a requirement, which severely limits analytical approaches. To make further progress, one can apply perturbation theory, if the deformations of the spherical shape are small.

\section{Perturbation theory}
\label{perturbation}
The active driving  will in general give rise to deformations of
the droplet which are counteracted by the homogeneous interfacial tension $\gamma$. 
 Following previous approaches for passive
droplets ~\cite{Rallison1984}, we now assume that this tension is large and  define a small parameter $\epsilon\ll 1$ by $\gamma=\hat{\gamma}/\epsilon$.  Regular perturbation theory  in $\epsilon$
will be used in the following only to lowest order, resulting in a
deformation of ${\cal {O}}(\epsilon)$. Yet it affects the
flow in order ${\cal {O}}(\epsilon^0)$, as we now show.

The deformed interface is described by  $\bm{r}(\theta)=r_s(\theta)\bm{e}_r$ with
\begin{equation}
r_s=1+\epsilon f(t,\theta)  
\end{equation}
in terms of the axisymmteric deformation $f(t,\theta)$. The isotropic contribution of $f$ is fixed by the volume constraint 
\begin{equation}
	2\pi\int_0^\pi(1+\epsilon f)^3 d(\cos\theta)=4\pi. 
\end{equation} Thus, to $O(\epsilon)$, the angular average of $f$ must vanish. We use the level set function
\begin{equation}\label{eq:interface}
H(\bm{r},t)=r-1-\epsilon f(t,\theta)=0, 
\end{equation}
so that the shape evolution Eq.(\ref{eq:kinetics}) becomes
\begin{equation}
\label{eq:perturbkinetic}
	\epsilon\frac{\partial f}{\partial t}=\bm{v}^\pm\cdot\bm{n}.
\end{equation}
The curvature term, which enters Eq.(\ref{eq:bc3}) is in general non-linear in $f$, but to lowest order in $\epsilon$ the normal unit vector is given by
\begin{align}
  \bm{n}=\bm{\nabla} H/|\bm{\nabla} H|&=\bm{\nabla} H+{\cal O}(\epsilon^2)\nonumber\\
  &=\bm{e}_r-\epsilon  \bm{\nabla}_s f+{\cal O}(\epsilon^2),
\end{align}
so that
\begin{equation}
\bm{\nabla}\cdot\bm{n}=2-\epsilon \nabla^2 f +{\cal O}(\epsilon^2)
\end{equation}
becomes linear in $f$.

The force balance  (Eq.\ref{eq:bc3}) to order $\mathcal{O}(\epsilon^0)$ then reads 
\begin{equation}
  \label{eq:bc5}
  \bm{t}^+-\bm{t}^-=\frac{\hat{\gamma}}{\epsilon}
  \big(2\bm{e}_r-2\epsilon\bm{\nabla} f-\epsilon(\nabla^2 f) \bm{e}_r
  \big)+\bm{t}_{act}.
\end{equation}
The Laplace pressure term  in Eq.(\ref{eq:bc5}) is proportional to the homogeneous surface tension and hence becomes large in our perturbative analysis. It is compensated by a jump of homogeneous pressure across the interface and it does not influence the flow. The other terms on the right hand side are all of order ${\cal O}(\epsilon^0)$. Thus, the corrections to the flow due to deformations of the droplet are of ${\cal O}(\epsilon^0)$, i.e. they will not vanish even for $\epsilon\to 0$ and have to be taken into account from the start. In the next section we obtain the flow field and the evolution of deformations, $f(\theta, t)$, of the driven droplet  from a complete analytical solution.

\section{ Analytical Solution}
\label{sec:analytical}

\subsection{Choice of basis}
In order to solve Stokes equations, it is convenient to start from vector spherical harmonics (VSH) $\bm{\psi}^s (s=1,2,3$), which diagonalize the surface Laplacian. For axially symmetric, achiral systems considered here we only need the two components with $s=1$ and $s=3$,
\begin{align}
\label{eq:psis1}
  \bm{\Psi}_l^1(\theta) &=P'_{l}(\cos\theta)\bm{e}_\theta + lP_{l}(\cos\theta) \bm{e}_r,\\
\label{eq:psis3}  
  \bm{\Psi}_l^3(\theta) &=P'_{l}(\cos\theta)\bm{e}_\theta - (l+1) P_{l}(\cos\theta) \bm{e}_r,
\end{align}
for $l=0,1,2, \cdots$. The $P_l(\cos\theta)$ denote Legendre polynomials and $P_l^{'}(\cos\theta) = dP_l(\cos\theta)/d\theta$. These VSH constitute a
$L^2$-complete and orthogonal set of axially symmetric, achiral vector fields on the unit sphere ( for the set including $s=2$ and for further details see {\cite{Kree2021, Kree2022}}).

A general solution of Stokes equations is constructed from the Ansatz
$\bm{u}_l^s(r,\theta)=g_l^s(r)\bm{\Psi}_l^s(\theta)$, leading to ordinary differential equations for $g_l^s(r)$, which do not couple different l.
Two complete sets of solutions of Stokes equations are found: one which is regular at the origin ($\bm{u}^{s<}$) and one which is regular at infinity ($\bm{u}^{s>}$). For a fixed value of l, they are explicitly given by:
\begin{align}
\label{eq:uin1}
\bm{u}^{1<} & = r^{l-1}\bm{\Psi}_{l}^{1} \\
\label{eq:uin3}
\bm{u}^{3<} & = \frac{r^{l+1}}{(2l+1)}\big[\bm{\Psi}_{l}^{1} + 2A_{l}\bm{\Psi}_{l}^{3}\big]\\
\label{eq:uout1}
\bm{u}^{1>} & = \frac{1}{r^{l+2}}\bm{\Psi}_{l}^{3} \\
\label{eq:uout3}
\bm{u}^{3>} & = -\frac{1}{(2l+1)r^{l}}\big[2B_l\bm{\Psi}_{l}^{1} + \bm{\Psi}_{l}^{3}\big].
\end{align}
with $ A_{l} = l/((2l+3)(l+1))$ and
$B_l=A_{-l-1}=-(l+1)/(l(2l-1))$. The normalization factors are chosen
for convenience. Note that these solutions are neither $L^2$
orthogonal nor normalized. We consider a fluid that rests at infinity
and write the general solution of Stokes equations for the interior ($\bm{v}^-$) and the exterior
($\bm{v}^+$) flows:
\begin{align}
  \label{eq:vminus}
  \bm{v}_l^-&=a_1\bm{u}^{1<}+a_3\bm{u}^{3<},\\
  \label{eq:vplus}
   \bm{v}_l^+&=c_1\bm{u}^{1>}+c_3\bm{u}^{3>}.
 \end{align}

To solve Stokes equation (\ref{eq:stokes}), subject to the incompressibility condition and the
boundary conditions Eqs.(\ref{eq:bc2}, \ref{eq:bc5}), we expand
the deformation $f$ in a Legendre series
$f(\theta,t)=\sum_{l} f_l(t) P_l(\cos\theta)$, and the drives in VHS
\begin{align}
  \bm{v}_{act}(\theta)&= \sum_l v_{a,l}\big((l+1)\bm{\Psi}_{l}^{1} +
                        l \bm{\Psi}_{l}^{3}\big)\\
  \bm{t}_{act}(\theta)&=\sum_l \big(t_{a,l}^1\bm{\Psi}_{l}^{1} +
                        t_{a,l}^3 \bm{\Psi}_{l}^{3}\big).
\end{align}
We specialise here to axisymmetric and achiral flow to keep the
subsequent discussion as simple as possible. The generalisation to non
axisymmetrc and chiral flow is straightforward.

 The $l=1$ terms of the flow determine the self-propulsion velocity
 $\bm{U}$. This flow does not deform the sphere, so the $l=1$
 component $f_1$ of deformations vanishes. The calculation of $\bm{U}$
 and the corresponding flow within our framework is outlined in
 Appendix {\ref{appendixB}}. In the main text, we focus on deforming
 drives with $l\geq 2$. In the next subsection, we discuss the
 explicit solution for the special case $l=2$ and only active tractions -- in order to demonstrate
 our line of aapproch. The case of general $l$ will be discussed
 subsequently with details of the calculation given in Appendix
 {\ref{appendixA}}.

\subsection{Complete solution for $l=2$}

We consider active tractions with fixed $l=2$, i.e.
$ \bm{t}_{act}=t_{a,2}^1\bm{\Psi}_2^1+t_{a,2}^3 \bm{\Psi}_2^3$, 
and discuss the flow, the relaxation of the deformation $f_2(t)$ and the stationary value $f_2^*$. As $l=2$ remains fixed throughout this subsection, we leave out the l-index in the following to lighten the notation.

From Eqs.(\ref{eq:uin1}-\ref{eq:uout3}), we get 

\begin{equation}
  \bm{u}^{1<}=r \bm{\Psi}^1,\qquad
  \bm{u}^{3<}=\frac{r^3}{5}\big( \bm{\Psi}^1+\frac{4}{21}
            \bm{\Psi}^3\big)\nonumber
            \end{equation}
and
\begin{equation}
  \bm{u}^{1>}=\frac{1}{r^4} \bm{\Psi}^3,\qquad
  \bm{u}^{3>}=\frac{1}{5r^2}\big( \bm{\Psi}^1- \bm{\Psi}^3\big).\nonumber
\end{equation}

We require that the interior $\bm{v}^-$ and  exterior flow $\bm{v}^+$, given by Eqs.(\ref{eq:vminus},\ref{eq:vplus}), are continuous across the droplet interface (no slip, only active tractions). Projecting the boundary condition of
 Eq. (\ref{eq:bc2})
onto $\bm{\Psi^1}$ and $\bm{\Psi}^3$, yields
 \begin{align}\label{c1}
   c_1 - c_3/5 & =4a_3/105\\
   \label{a1}
   a_1 + a_3/5 & =c_3/5.  
 \end{align}

 The balance of  tractions, Eq. (\ref{eq:bc5}), at the droplet interface requires computation of the viscous tractions
\begin{equation}\label{tractions}
\bm{t}[\bm{v}]=  -p\bm{e}_{r}  + 2\eta \big(\bm{e}_{r}\cdot\nabla\big)\bm{v} + \eta\bm{e}_{r}\times \big(\nabla\times\bm{v} \big),
\end{equation}
valid on a sphere $r=const$ in a fluid of viscosity $\eta$, together with Stokes equations. The resulting expressions take on the form ($r=1$):
 \begin{align}
\label{eq:tranfang}
\bm{t}[\bm{u}^{1<}] &= 2\eta^-\bm{\psi}^1\\
\bm{t}[\bm{u}^{3<}] & = \frac{2\eta^-}{5}
\Big[\bm{\psi}^1 + \frac{19}{21} \bm{\psi}^3\Big]\\
\bm{t}[\bm{u}^{1>}] & = -8\eta^+\bm{\psi}^3\\
\bm{t}[\bm{u}^{3>}] & = \frac{2\eta^+}{5}
\Big[4\bm{\psi}^3 - \frac{3}{2}\bm{\psi}^1 \Big].
\label{eq:trende}
\end{align}
Substituting these expressions into the boundary condition (Eq.\ref{eq:bc5})
yields 
\begin{align}\label{t1}
  2\eta^-(a_1+a_3/5)+3\eta^+c_3/5&=-t_a^1\\
  \label{t2}
 8\eta^+(c_1-c_3/5) + 2\eta^- 19a_3/105 &=2{\hat{\gamma}}f-t_a^3. 
 \end{align}
Thus we  obtain a system
 of 4 linear equations to determine the yet unknown coefficients   $(a_1,a_3,c_1,c_3)$. For $l=2$, these equations are easily solved:
we substitute the expression (Eq.\ref{a1}) for $a_1=(a_3+c_3)/5$ into Eq.(\ref{t1})
and the expression (Eq.\ref{c1}) for $c_1=(4a_3+21c_3)/105$ into Eq.(\ref{t2}) to obtain
  \begin{align}
   c_3&=-5 t_a^1/(3\eta^++2\eta^-)\\
a_3&=105({\hat{\gamma}}f-t_a^3/2)/(16\eta^++19\eta^-). 
 \end{align}
 Finally, we re-insert these expressions into Eqs.(\ref{c1} and \ref{a1}). Thereby the flow fields for a general deformation $f$ have been calculated.

 The time evolution of the interface is given by  Eq.(\ref{eq:perturbkinetic}), and takes on the form 
 \begin{equation}\label{dyn_int_explicit}
   \epsilon\frac{\partial f}{\partial t}= \bm{v}^\pm\cdot\bm{e}_r=-\frac{2t_a^1}{3\eta^++2\eta^-}
   -6\frac{2{\hat{\gamma}}f-t_a^3}{(16\eta^++19\eta^-)}.
 \end{equation}
 The stationary deformation $f^*$ is determined from the condition
 $\bm{v}\cdot \bm{e}_r=0$, and depends upon the values of $t_{a}^1$
 and $t_{a}^3$.  Given a specific driving, the
 coefficients $t_{a}^1$ and $t_{a}^3$ are obtained by projecting
 $\bm{t}_{act}$ onto $\bm{\Psi}^1 $ and $\bm{\Psi}^3$.  As a well
 studied example system we consider Marangoni flows on the surface of
 the droplet. 
 A nonuniform surface tension
 $\gamma(\theta)=\hat{\gamma}/{\epsilon} +\delta\gamma P_2(\theta)$ gives rise to active tractions: $\bm{t}_{act}=-\delta\gamma (\bm{\Psi}^1+4\bm{\Psi}^3)/5$.
These in turn deform the droplet, such that its stationary state is characterized by the following deformation amplitude
\begin{equation}
\label{eq:stationaryl2}
   f^*= -\frac{\delta\gamma(4\eta^++\eta^-)}{6(3\eta^++2\eta^-)}.
 \end{equation}
 
 To show that the stationary deformation $f^*$, which is a zero of the right hand side of Eq.(\ref{dyn_int_explicit}), corresponds to a stable shape of the droplet, consider time dependent fluctuations, $f(t)=f^*+\delta f(t)$, away from the stationary value $f^*$. Substituting this ansatz into Eq.(\ref{dyn_int_explicit}) yields an exponential decay of $\delta f(t)$
 with relaxation time 
 \begin{equation}
   \tau=\frac{\epsilon}{{\hat{\gamma}}}\frac{16\eta^++19\eta^-}{12}
\end{equation}
We conclude that the droplet configuration, described by $f^*$, is stable and that deviations from the stationary state relax quickly due to a large surface tension.


The flow field far away from the droplet falls off as $r^{-2}$ and is purely radial
\begin{equation}\label{eq:farfield}
  \bm{v}_2^+ {\rightsquigarrow}\frac{\gamma_2}{r^2(3\eta^++2\eta^-)}P_2(\cos\theta)\bm{e}_r
  \end{equation}
  The deformation and flow field are shown in Fig.\ref{fig:flowdeform}(a) and (d). To make the small
deformations clearly visible, they are blown up by a factor $1/\epsilon$.
Note that the flow has been computed only to order
  ${\cal O}(\epsilon^0)$ and thus fulfills the condition
  $\bm{v}\cdot\bm{n}=0$  on the undeformed spherical interface.

\subsection{General solution}
The method introduced in the previous subsection can be generalized to any fixed $l$ and to all kind of drives, $\bm{v}_{act}$ and $\bm{t}_{act}$.  The linear system for the coefficients $a_1,a_3,c_1,c_3$  and its solution are derived explicitly in Appendix\ref{appendixA} and summarized in Table \ref{tab:coeff}. The active drives appear as inhomogeneities in the linear system, and therefore the solutions take on the form of a sum of terms, each proportional to one special inhomogeneity, $v_a^s, t_a^s, s=1,3 $. 
The terms arising from the deformation $f_l$ in Eq.(\ref{eq:bc5}) may be added to this list and are referred to as $t_f^s$. They take on the forms given in Eq. (\ref{eq:appendtfa}) and Eq.(\ref{eq:appendtfb}).
The solution for a special drive can now be obtained as a linear combination of the  corresponding contributions in table \ref{tab:coeff}.

The solutions $\bm{v}^\pm$ are obtained as linear combinations of VSH,
but can easily be re-expressed in terms of radial and tangential
components,  which are easier to visualize. Using Eqs.(\ref{eq:psis1}, \ref{eq:psis3}), a fixed $l$
part of a vector field,
$\bm{v}_l(\theta)=v_{l}^1\bm{\Psi^1} + v_l^3\bm{\Psi}^3$, can be
rewritten as
$\bm{v}=v_{l,r}P_l(\cos\theta)\bm{e}_r+
v_{l,\theta}P'(\cos\theta)\bm{e}_\theta$, with
$v_{l,r} = l v_l^1-(l+1)v_l^3$ and $v_{l,\theta} = v_l^1 + v_l^3$.
The
interior flow is given by
$\bm{v}^-(r,\theta)=v^-_r(r)P_l(\cos\theta)\bm{e}_r +
v^-_\theta(r)P'(\cos\theta)\bm{e}_\theta$ with
\begin{align}
\label{eq:vminusr}
	v^-_r(r)  = & l r^{l-1}\left(a_1+\frac{a_3r^2}{2l+3}\right)\\
	v^-_\theta(r) = & r^{l-1}\left(a_1 + \frac{a_3r^2(l+3)}{(l+1)(2l+3)}\right),
\end{align}
and the exterior flow is characterized analogously by
\begin{align}
\label{eq:vplusr}
	v^+_r(r)  = & \frac{(l+1)}{r^{l+2}}\left(-c_1 + \frac{c_3r^2}{2l-1}\right)\\
	v^+_\theta(r) = & \frac{1}{r^{l+2}}\left(c_1 - \frac{c_3r^2(l-2)}{l(2l-1)}\right).
\end{align}
Note that only for $l=2$ the leading order term of the exterior flow is purely radial, as found in Eq.(\ref{eq:farfield}).

The radial component $v_r$ of the flow velocities on the spherical interface, which determine the shape evolution, are given in Table \ref{tab:vr}. From $v_r$, the stationary shape $f^*$ and the relaxation time for fluctuations away from this shape can be obtained.

\section{Stationary shapes}
\label{sect:Comparison}
Stationary shapes and the corresponding flow fields can be obtained from the general solution, but
the stationarity conditions $v_r^+ = v_r^- =0$ and the tangential part of Eq. (\ref{eq:bc2}) already strongly constrain the functional form of the flow and simplify the calculations, if they are applied from the start. From Eq.(\ref{eq:vminusr}) we get  $a_3=-(2l+3)a_1$   and from Eq.(\ref{eq:vplusr}) $c_3=(2l-1)c_1$ is found. 
Furthermore, the two remaining free constants $a_1$ and $c_1$ are related by
\begin{equation}
\label{eq:stationarytangential}
	\frac{a_1}{l+1} + \frac{c_1}{l}=\frac{v^a_{l,\theta}}{2},
\end{equation}
so that only one constant per l-component has to be calculated from force balance, together with the stationary deformation $f^*$. The two force balance equations simplify considerably for stationary shapes (see  (see Eqs.(\ref{eq:trac1}, \ref{eq:trac2}))) and lead to 
\begin{align}
\label{eq:stationaryf}	
		D\hat{\gamma}f^* = &-3\eta^-(2l+1)v_{a,\theta} -(2l+1)(1+\lambda)t_{a,r}\\ \nonumber
		& -3(l+(l+1)\lambda)t_{a,\theta}
\end{align}
  
with
\begin{equation}
	D=(1+\lambda)l(l-1)(2l+5))-6\lambda
\end{equation}
As a special case,  Eq.(\ref{eq:stationaryl2}) of $l=2$ interfacial tractions is recovered here, using $t_{a,r}=2\delta\gamma$ and $t_{a,\theta}=-\delta\gamma$.
    
\begin{figure*}[t]
\begin{subfloat}{(a)}
\centering
\includegraphics[width=0.21\textwidth]{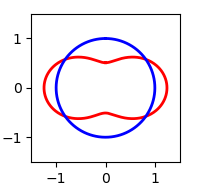}
\end{subfloat}
\hspace*{\fill}
\begin{subfloat}{(b)}
\centering
\includegraphics[width=0.2\textwidth]{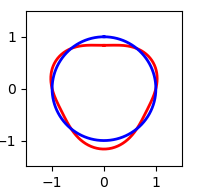}
\end{subfloat}
\hspace*{\fill}
\begin{subfloat}{(c)}
\centering
\includegraphics[width=0.2\textwidth]{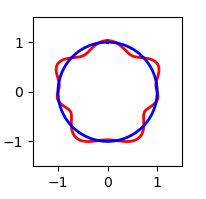}
\end{subfloat}
\hspace*{\fill}
\\
\centering
\begin{subfloat}{(d)}
\centering
\includegraphics[width=0.3\textwidth]{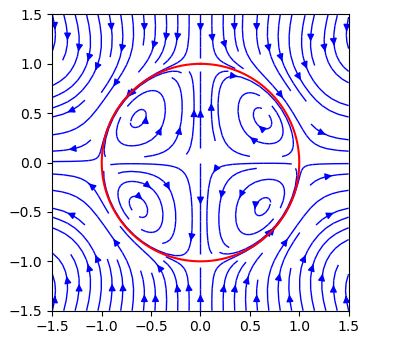}
\end{subfloat}
\begin{subfloat}{(e)}
\centering
\includegraphics[width=0.3\textwidth]{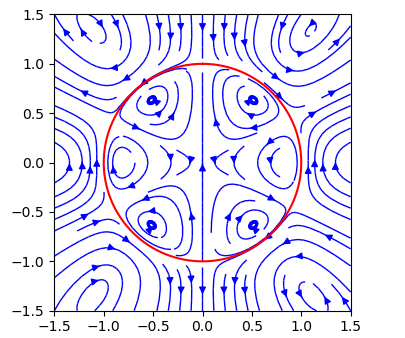}
\end{subfloat}
\begin{subfloat}{(f)}
\centering
\includegraphics[width=0.3\textwidth]{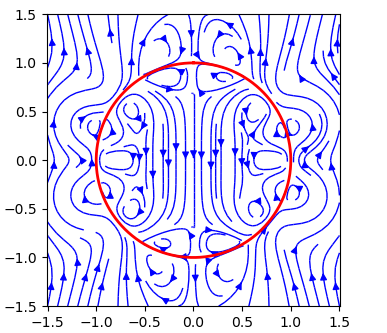}
\end{subfloat}
\caption{\emph{First row}: $x-z$ plane cut of deformed droplets. Deformations are due to inhomogeneous surface tensions with l=2 (a), l=3 (b) and the Legendre series of $\sin(10z)$ up to $l=20$ (c). The deformations are blown up by a factor $1/\epsilon$ to make them clearly visible. \emph{Second row}: (d)-(f) show the corresponding flow fields. } 
\label{fig:flowdeform}
\end{figure*}

Some examples of stationary deformations and the corresponding flows are shown in Fig.\ref{fig:flowdeform}. Panels (a,d) and (b,e) correspond to single l-components (l=2,3) of $\delta\gamma$. 
An example of a more complex flow, involving all $l$,
is generated by
an inhomogeneous
surface tension $\delta\gamma=\sin(10\cos{\theta})$. 
To apply our framework, we expand $\delta\gamma$ into a Legendre series and  take into account terms up to order $l=20$. This reproduces
$\delta\gamma$ with a typical absolute error smaller than $5\times 10^{-4}$ (and a
maximal error smaller than $\times10^{-3}$ at the boundaries $z=\pm 1$).
Fig. \ref{fig:flowdeform} (c) shows the deformation and (f)  the flow field generated by these tractions. Note that the Legendre series contains an $l=1$ term, which leads to self-propulsion and requires the calculation of a flow field as described in Appendix\ref{appendixB}.




\subsection{Non-deformation boundary conditions}
In a regime of small capillary number
$Ca: =\epsilon |\bm{t}_{act}|/\hat{\gamma}$ several authors
\cite{Lauga2017, Schmitt2016} discard all effects of deformations and
require $\bm{v}^{\pm}\cdot\bm{n}=0$. Together with 
continuity of tangential flow and the
balance of tangential tractions, this reduces the solution of the
boundary value problem to a $2\times 2$ linear system, consisting of
Eq.(\ref{eq:stationarytangential}) and Eq.(\ref{eq:trac2}) with
$f^*=0$.

Intuitively it may seem plausible that an infinitely large interfacial tension will keep the droplet shape spherical. It should be noted, however, that even in the limit $\epsilon, Ca\to 0$, the effects of $f^*$ on the flow field do not vanish.  The non-deformation boundary conditions therefore require extra, non-physical radial tractions $\Delta t_r$ to fulfill the complete traction balance and suppress the deformation. These tractions are calculated by inserting the non-deformation solution into the radial balance Eq.(\ref{eq:trac1}), and are given by
\begin{equation}
\Delta t_r = -\frac{3\eta^-}{1+\lambda}v_{a,\theta} -
 \frac{l+(l+1)\lambda}{1+\lambda} \frac{t_{a,\theta}}{2l+1} - t_{a,r}.	
\end{equation}
It is unclear, how these radial tractions should arise, in particular so, because they depend on the activity of the droplet.

It is instructive to study the error induced by replacing the correct
boundary conditions for $\epsilon \to 0$ with non-deformation
conditions. This error should be large if large radial tractions
appear. As an extreme case, a drive consisting only of radial
tractions generates a non-vanishing flow field of
$\mathcal{O}(\epsilon^0)$, whereas non-deformation boundary conditions
lead to no flow at all.

More generally, we can quantify the error in the exterior flow for a given fixed l  by $\Delta c=c_1-c_{0}$, where $c_{0}$ is the
$c_1$-coefficient obtained from non-deformation boundary conditions. Due
to Eq.(\ref{eq:stationarytangential}) this also determines the error
in the internal flow, $\Delta a/(l+1)=-\Delta c/l$. From
Eq.(\ref{Appeq:stationaryf}) one reads
off that
\begin{equation}
	\Delta c= \frac{l\hat{\gamma} f^*}{(\eta^+ + \eta^-)(2l+1)}.
      \end{equation}
      Substituting the result for $f^*$ from Eq.(\ref{eq:stationaryf})
      we obtain the relative errors $\Delta c/c_0$  for inhomogeneous surface tension shown in  Fig.{\ref{fig:differences}(a) and similarly for  active slip, shown in Fig.{\ref{fig:differences}(b). The error can be as large as $25\%$
 for $l=2$ and decreases rapidly for growing l. 

\begin{figure}[t]
\begin{subfloat}{(a)}	
\centering
\includegraphics[width=0.9\columnwidth]{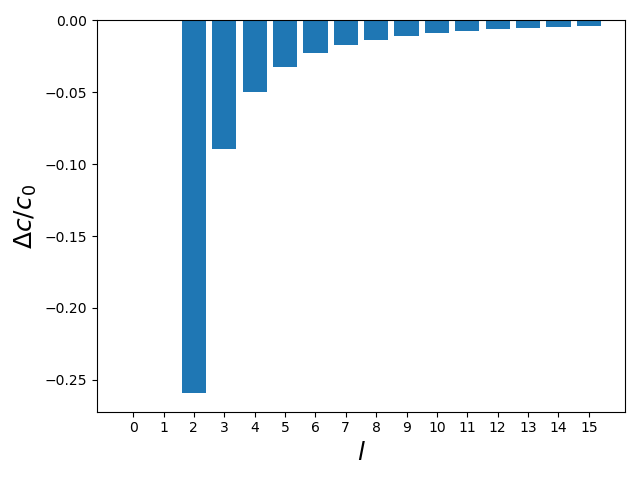}
\end{subfloat}\\
\begin{subfloat}{(b)}
\centering
\includegraphics[width=0.9\columnwidth]{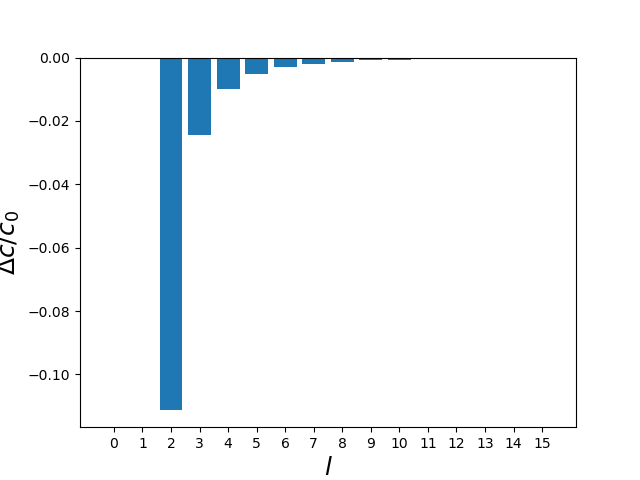}
\end{subfloat}
\caption{Normalized errors in exterior flow between correct boundary conditions for $\epsilon\to 0$ and non-deformation boundary conditions vs. $l$: (a) for inhomogeneous surface tension and b) for active slip velocities.}
\label{fig:differences}
\end{figure}

\section{Discussion and outlook}

A consistent approach to the flow fields inside and outside a liquid
droplet, which is driven by a generic activity mechanism,
has to include deformations of the droplet.  For a large homogeneous surface tension $\gamma$, the deformations are small of order
$\epsilon \propto 1/\gamma$. However, these small deformations have a {\it
  finite} effect on the flow, even in the limit $\epsilon\to 0$.
This surprising result can be traced back to Eqs.(\ref{eq:bc3},\ref{eq:bc5}).
Small deformations, away from the spherical shape, correspond to a
non-uniform curvature of the surface, i.e.  $\bm{n}\neq\bm{e}_r$. In
the resulting tractions due to the curvature term,
$\bm{t}=\gamma\bm{n}(\bm{\nabla}\cdot\bm{n})$, the large homogeneous surface tension is multiplied by the small
nonuniform curvature, so that the overall tractions due to deformations of the droplet stay finite  for $\epsilon\to 0$. 

We have calculated the flow fields and the dynamics of deformations using a versatile method based on special solutions of the Stokes equation, derived from VSH. For all possible drives due to active slip velocities and/or active tractions we find locally stable stationary deformations. 

In order to achieve an easy comprehensibility of our presentation, we
only considered the axially symmetric, achiral case, but the extension
to non-axisymmetric systems including swirling flow is
straightforward. The necessary system of solutions of Stokes equations
for the most general case can be found in \cite{Kree2021, Kree2022}.

We compared our results to flows computed from non-deformation boundary conditions. The approximation made in this set of conditions does not affect the velocity of self-propulsion, as the $l=1$ flow generated by propulsion does not deform the spherical droplet. Flow components with $l$ between 2 and 10, however, differ significantly and may lead to discrepancies in the range 25\%- 5\% of the flow velocities.

So far we have looked at the lowest order perturbation theory in
surface tension.  A natural next step is the extension to higher
orders. Already the next order will give rise to curvature terms, which are nonlinear in the deformation $f$, implying a
coupling of different $l$. Thereby even drives with $l\geq 2$ can give rise to self propulsion due to shape deformations of the
droplet. 

Another possible application of our approach are controlled time dependent deformations $f(\theta,t)$ of the droplet, which provides an analytically tractable model of amoeboid propulsion.

\appendix
\section{Self-propulsion}
\label{appendixB}
Here we show how to calculate the self-propulsion velocity $\bm{U}$
and the flow resulting from $l=1$ components of the drives using the
general formalism, as explained above. Due to axial symmetry the
propulsion velocity points in z-direction and thus
$\bm{U}=U\bm{e}_z=U\bm{\Psi_1^1}$. In contrast to the main part of the paper, we work here in the frame (CMF), where the external flow does not decay to zero for $r\to \infty$.  In order to use the expansion of Eq.(\ref{eq:vplus}), we
first have to decompose the exterior flow as
$\bm{w}^+=\bm{v}^+ - \bm{U}$ and then expand $\bm{v}^+$
as in Eq.(\ref{eq:vplus}).

The $l=1$ flow does not change the spherical shape of the droplet,
which implies $f_1(t)=f^*=0$ in the CMF, and consequently the radial
velocity components on the interface vanish,
i.e. $\bm{v}^-\cdot\bm{e}_r=0$ and $\bm{w}^+\cdot\bm{e}_r=0$, implying
$\bm{v}^+\cdot\bm{e}_r=UP_1(\cos\theta)$. From
Eq.(\ref{eq:radialcomponent}) this gives $a_3=-5a_1$ and
$c_3=c_1+U/2$. Inserting these relations into Eqs. (\ref{eq:appendva},
\ref{eq:appendvb}) gives two equations, but they are linearly
dependent, so that the 3 remaining unknowns ($a_1,c_1, U$) are
determined by the equation
\begin{equation}
\label{eq:selfpropVelocity}
	2c_1-U/2 +a_1=v_{\theta}
\end{equation} 
and the two Eqs.(\ref{eq:tracboundary1}, \ref{eq:tracboundary2}). In particular, Eq.(\ref{eq:tracboundary1}) determines $c_3=-t_a^1/2\eta^+$. But $t_{a}^1$ has to vanish, because the total force $\bm{F}$ of self-propelling tractions only stems from the $l=1$ components, so that
\begin{equation}
	\bm{F}=\int_0^\pi (t_{a}^1 \bm{\Psi_1^1} + t_{a}^3 \bm{\Psi_1^3}) d\theta= 2 t_{a}^1 \bm{e}_z. 
      \end{equation}     
 For an active slip velocity, the condition $c_3=0$ is always fulfilled, and is equivalent to the vanishing of total viscous force of the flow, i.e. the integral over tractions on the infinite sphere.   
As a consequence, $c_1=-U/2$ and $a=v_\theta + 3U/2$. Finally we can insert these results into Eq.(\ref{eq:tracboundary2}) to determine $U$, which becomes
\begin{equation}
\label{eq:propulsion}
	U=\frac{2t_a^3/3 - 2\eta^- v_\theta}{2\eta^+ + 3\eta^-}
      \end{equation}
      
      Note that force free tractions require both, a tangential
      component $t_{a,\theta}=t_a^3$ and a radial component
      $t_{a,r}=-2t_a^3$. However, the force balance equations for the
      radial and tangential component are redundant, so that
      non-deformation boundary conditions give the same result as
      Eq.(\ref{eq:propulsion}).  For inhomogeneous interfacial
      tensions, $t_a^3=-\delta\gamma_1$, 
      Eq.(\ref{eq:propulsion}) reproduces the results from
      \cite{Levan1976}. General tractions were previously analysed in
      \cite{Kree_2017} with however incorrect results for
      active slip velocites.
 
\section{General analytical solution}
\label{appendixA}
In this appendix we present details of the analytical solution for general $l$.
Flow and shape evolution of an active droplet, driven by axially symmetric active slip velocities and/or active tractions are derived in lowest order of perturbation theory in $1/\gamma$. The boundary conditions Eqs.(\ref{eq:bc2},\ref{eq:bc5}) are imposed as well as the kinematic Eq.(\ref{eq:perturbkinetic}). 
 
 To lowest order in
perturbation theory, the boundary value problem is linear and terms with different $l$ remain uncoupled.
Therefore we only need to consider drivings with a single fixed $l$,
which lead to deformations given by $f_l P_l(\cos\theta)$.
The fixed index $l$ will be left out in the following.

We insert the expansions, Eq.(\ref{eq:vminus}) and Eq.(\ref{eq:vplus})
into the condition Eq.(\ref{eq:bc2}) and project onto $\Psi^1$ and $\Psi^3$ to get the following two equations

\begin{align}
\label{eq:appendva}
	a_1 + \frac{a_3}{2l+1} &= -\frac{2c_3B_l}{2l+1} -v_a^1\\
\label{eq:appendvb}
	c_1-\frac{c_3}{2l+1} & =\frac{2a_3A_l}{2l+1} +v_a^3.
\end{align}

The boundary conditions Eqs.(\ref{eq:bc5}) contain the tractions on the spherical interface. They are obtained from Eq.(\ref{tractions})
\begin{align}
\label{eq:tranfang}
\bm{t}[\bm{u}_l^{1<}] &= 2\eta^-(l-1)\bm{\psi}_{l}^1\\
\bm{t}[\bm{u}_l^{3<}] & = \frac{2\eta^-}{(2l+1)}
\Big[(l-1)\bm{\psi}_{l}^1 + (1- A_l) \bm{\psi}_{l}^3\Big]
\end{align}
and 
\begin{align}
\bm{t}[\bm{u}_l^{1>}] & = -2\eta^+(l+2)\,\bm{\psi}_{l}^3\\
\bm{t}[\bm{u}_l^{3>}] & = -\frac{2\eta^+}{(2l+1)}
\Big[(1- B_l) \bm{\psi}_{l}^1 - (l+2)\bm{\psi}_{l}^3  \Big].
\label{eq:trende}
\end{align}
Using these expression in Eq.(\ref{eq:bc5}), we get the two equations
\begin{align}
\label{eq:tracboundary1}
	2\eta^-(l-1)\left(a_1 + \frac{a_3}{2l+1} \right) + \frac{2\eta^+}{2l+1}(1-B_l)c_3 &= -T_a^1\\
	2\eta^+(l+2)\left(c_1-\frac{c_3}{2l+1}\right) + \frac{2\eta^-}{2l+1}(1-A_l)a_3 &=-T_a^3.
\label{eq:tracboundary2}
\end{align}
The tractions $T_a^s$ on the right hand side are composed of active tractions $t_a^s$ and the tractions arising from deformations, $T_a^s=t_a^s + t_f^s$, with 
\begin{align}
\label{eq:appendtfa}
	t_{f}^1 &= \frac{(l+1)(l-2)}{2l+1}\hat{\gamma}f\\
	\label{eq:appendtfb}
	t_{f}^3 & = -\frac{l(l+3)}{2l+1}\hat{\gamma}f.
\end{align}
Now we can insert Eq.(\ref{eq:appendva}) into Eq.(\ref{eq:tracboundary1}), and Eq. (\ref{eq:appendvb}) in Eq.(\ref{eq:tracboundary2}) to obtain 
$a_3$ and $c_3$. These can be re-inserted into Eqs. (\ref{eq:appendva}, \ref{eq:appendvb}) to completely determine the flow field. The contributions to $a_s, c_s$ are listed in table \ref{tab:coeff}. They are obtained from the table by multiplying each cell with the entry in the rightmost column of the same row, where we have used the abbreviated expressions
\begin{align}
	M & =2l^2(\lambda +1) - (2\lambda - 1)\\
	N & =2l(l+2)(\lambda +1) +3\lambda.
\end{align}

Once the flow field is determined, we get the evolution of deformation $f$ from $\bm{v}^\pm(r=1)\cdot\bm{e}_r=v_r$. In terms of $c_1$ and $c_3$ it is given by 
\begin{equation}
\label{eq:radialcomponent}
	v_r=(l+1)\left(\frac{c_3}{2l-1}-c_1\right)=l\left(a_1+ \frac{a_3}{2l+3}\right).
\end{equation}
 The corresponding contributions are listed in table \ref{tab:vr}.

 For stationary flows, Eq.(\ref{eq:radialcomponent}) predicts $v_r=0$
 and the radial and tangential components of the corresponding
 tractions ${t}[\bm{v}^\pm]$ are obtained from Eqs (\ref{eq:tranfang}
 - \ref{eq:trende}). With Eq. (\ref{eq:psis1}, \ref{eq:psis3}) and
 Eqs.(\ref{eq:vminus}, \ref{eq:vplus}) they take on the form
 \begin{align}
\bm{e}_r\cdot\bm{t}[\bm{v}^-]	&= 6\eta^- a_1 P_l\\
\bm{e}_r\cdot\bm{t}[\bm{v}^+]	&= 6 \eta^+ c_1P_l\\
\bm{e}_\theta\cdot\bm{t}[\bm{v}^-]	&= -2\eta^- a_1\frac{2l+1}{l+1}P'_l\\
\bm{e}_\theta\cdot\bm{t}[\bm{v}^+]	&= -2\eta^+ c_1 \frac{2l+1}{l}P'_l
\end{align}
Using these expressions in the boundary condition Eq. (\ref{eq:bc5}), one gets 
\begin{align}
\label{eq:trac1}
  \eta^+c_1-\eta^- a_1 &=\frac{1}{6}[l(l+1)\hat{\gamma} f^* + t_{r}] \\
\label{eq:trac2}
  \frac{\eta^+ c_1}{l} - \frac{\eta^- a_1}{l+1} & =
                                              -\frac{t_{\theta}- 2\hat{\gamma} f^*}{2(2l+1)}.
\end{align}
Together with Eq.(\ref{eq:stationarytangential}), they are easily solved and lead to Eq.(\ref{eq:stationaryf}) and to 
\begin{equation}
  \label{Appeq:stationaryf}
	\frac{\eta^+ + \eta^-}{l} c_1= -\frac{1}{2}\left(v_{a,\theta} + \frac{t_{a,\theta} - 2 \hat{\gamma}f^*}{2l+1}\right)
\end{equation} 

The analogous calculations for non-deformation boundary conditions imply  $c_0=c(f^*=0)$.

\begin{widetext}
\begin{table}
$$
\begin{array}{|c|c|c|c||c|}
\hline
a_1 & a_3 & c_1 & c_3 &\\
\hline\hline
 -(l+1)& 0 & -l(2l-1)/2 & -l(2l-1)(2l+1)/2 & \cdot T_a^1/M \\
\hline
(l+1)(l+3)/2 & (l+1)(2l+1)(2l+3)/2 &-l &0 & \cdot T_a^3/N  \\	
\hline
-(2l^2+1) & 0 & \lambda l(l-1)(2l-1) &\lambda (l-1)(2l+1)(2l-1) & 
\cdot v_a^1/M  \\
\hline
(l+1)(2l+1)(2l+3) &-(l+1)(l+2)(2l+1)(2l+3) &\lambda(l^2+4l+3) & 0 & \cdot v_a^3/N \\
\hline
\end{array}
$$
\caption{\label{tab:coeff}}
\end{table}
\end{widetext}

\begin{table}
$$
\begin{array}{|c|c|}
\hline
v_r & \\
\hline
\hline
 -l(l+1)  & \cdot T_a^1/M\\
\hline
 l(l+1) & \cdot T_a^3/N \\	
\hline
2\lambda l (l-1)(l+1) & \cdot v_a^1/M\\
\hline
-\lambda(l+1)(2l^2 + 4l +3) & \cdot v_a^3/N\\
\hline
\end{array}
$$
\caption{\label{tab:vr}}
\end{table}

\textbf{Author's contributions:}\\
Both authors contributed equally to this work\\

\textbf{Data availability:}\\
Data available in article

\bibliography{squirmer.bib}


\pagebreak

\end{document}